# Dual Plane Imaging

*I. R. Parry[1] and A. M. Moore[2]*

[1]Cambridge University, Institute of Astronomy, Madingley Rd, Cambridge, CB3 0JA, UK, E-mail: irp@ast.cam.ac.uk

[2]California Institute of Technology, 1200 E. California Blvd. Pasadena, CA 91125, USA, E-mail: amoore@astro.caltech.edu

**Abstract:** We outline a technique called Dual Plane Imaging which should significantly improve images which would otherwise be blurred due to atmospheric turbulence. The technique involves capturing all the spatial, directional and temporal information about the arriving photons and processing the data afterwards to produce the sharpened images. The technique has particular relevance for imaging at around 400-1000nm on extremely large telescopes (ELTs).

## 1. INTRODUCTION

The traditional technique for correcting the wavefront errors introduced by the Earth's atmosphere is to use an adaptive-optics (AO) system which measures the wavefront errors in real time and applies an optical correction via some active optical device. A guide star (natural or laser) is needed to measure the wavefront error and the correction has to be applied quickly before the atmospheric conditions change substantially. The degree of correction depends on the number of independent actuators across the wavefront in the active optical device. This technique works well on present day 4-10m telescopes in the near-IR. The big challenge is to make AO systems that work well on 20-100m class ELTs at optical (~500nm) wavelengths.

In this paper we present an alternative approach to correcting images for atmospheric wavefront errors: Dual Plane Imaging (DPI). The name comes from the fact that for the field of view (FOV) of interest the light intensity distribution is recorded in full by a photon counting detector for both the image (sky) plane and the pupil (atmosphere or primary mirror) plane. Furthermore, this information is recorded on a timescale corresponding to that of the atmospheric turbulence. By processing this data the corrected sky images can be generated. This processing does not have to be in real time.

If there were no wavefront errors due to the atmosphere, the light from a guide star would form an Airy disk in the focal plane and a pupil image formed with this light would have a uniform illumination. By comparing this to what is actually observed in real pupil images it is possible to generate a transformation algorithm that moves the photons in the real data to where they would be in the ideal case of no wavefront errors. This algorithm can be applied to the whole data set to create a corrected image of the whole FOV. Note that this can only be done by measuring the path taken by every photon. Hence the need to record the image in both the pupil and image planes (i.e. record a pupil image for each spatial point in the FOV).

Placing a detector directly in the image plane and taking a long exposure is the traditional way of taking an image. In this case only the spatial information is recorded. The so called "lucky imaging" technique[1] takes this one step further by recording the temporal information as well. High definition images are obtained by selecting images captured during moments of excellent seeing. Unfortunately, the frequency with which such "lucky" moments occur reduces drastically as telescope aperture increases. The aperture limit for this technique is approximately around 2-2.5m. Our DPI technique goes one step further again by capturing the *directional* information about the arriving photons as well as their temporal and spatial distributions. We expect that the DPI technique will offer particular advantages for large (20-100m) telescopes.

## 2. THE OPTICAL SYSTEM

A simple implementation of the DPI technique consists of a lens array in the FOV which generates an array of pupil images on a detector (or array of detectors) placed behind it. Some fore-optics will be required to change the physical scale of the FOV to match the detector pixel size. See figure 1.

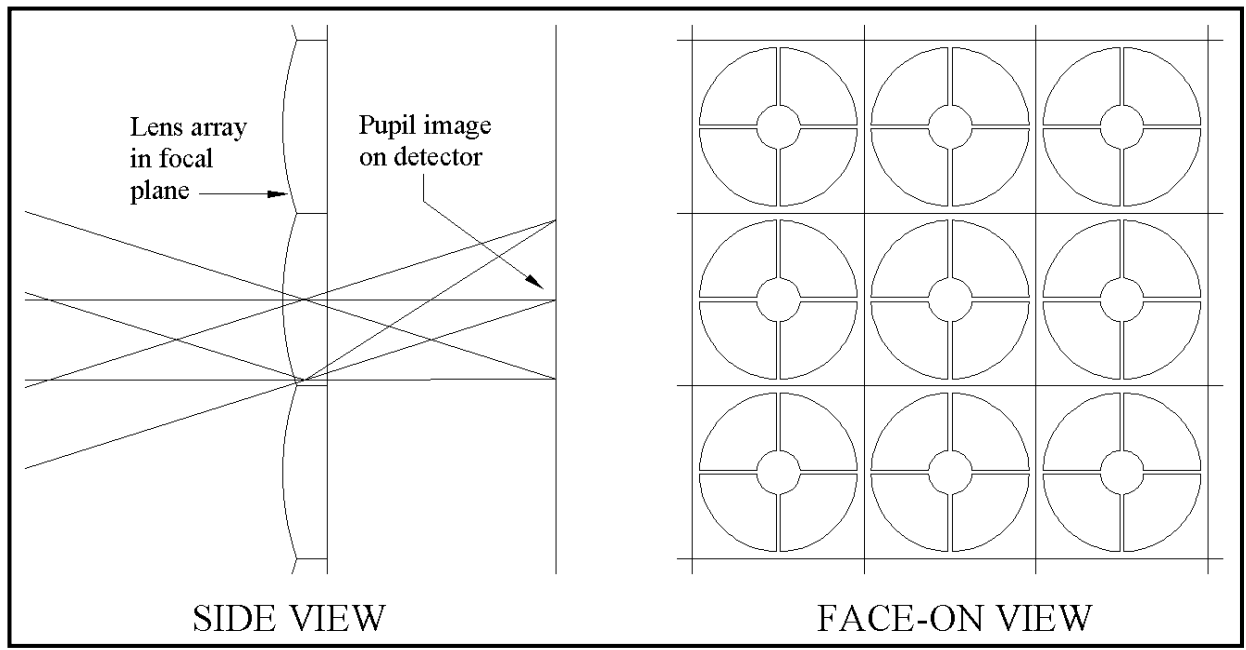

**Figure 1:** Part of the optical system for DPI. An array of lenses (in this case a square array) is placed at the focal plane. The figure only shows a 3×3 part of the full FOV. Each lens samples one pixel in the final image. Each lens also creates a pupil image on the detector which is placed behind the lens array. The position of a photon event in a pupil image gives the direction from which it came. The position it arrived at in the focal plane is determined by which pupil image it is seen in. The side view shows the optical rays. The face-on view shows the array of square lenses and the array of pupil images. The fore-optics required to appropriately scale the FOV to match the detector size is not shown.

### 3. THE DATA CAPTURE SYSTEM:

To capture the required data we need a very large number of detector pixels and each pixel has to be essentially noise free, i.e. a photon counting device. Such detectors do currently exist. For example, the L3CCD detectors from e2V[2] are photon-counting CCDs. For each pixel in the sky image we have to record a pupil image of ~10×10 pixels and we have to do this every 10ms (100Hz). Clearly, this is an enormous data rate and requires a very fast data capture system with a very large data storage capacity. For example, if we use 10 detectors of 1k×1k pixels each then we would record ~76 Terabytes of data per night.

### 4. THE TRANSFORMATION ALGORITHM

A big advantage of the DPI technique is that the applied correction can be intelligent and optimised. This is not true for classical AO systems where the algorithm that servos the adaptive optical device has just one chance to get it right and little is known at the end of the exposure as to the actual corrections made to the image over extremely short time scales.

With DPI many algorithms can be used and the confidence in the correction applied can be quantified. The technique should have very good photometric accuracy. For periods of very bad atmospheric behavior the data can simply be rejected rather than allowed to compromise the higher quality data (although it is hoped that the amount of data rejected will be much less than for the "lucky imaging" technique). The algorithm does not just have to apply to a single time sample of data so that temporal atmospheric correlations can be fully modeled and exploited. Many stars in the FOV can be used as "guide stars" (i.e. stars used to correct the image). With an iterative transformation algorithm more stars can be used as the algorithm converges to the optimized solution. Also the field dependence of the correction can be evaluated and applied by using several stars. The DPI technique collects at least as many photons from the guide stars as an equivalent AO system. This, combined with the technique's computational freedom, should lead to useful correction with fainter natural guide stars and therefore better sky coverage than for an AO system.

## 5. DIFFRACTION EFFECTS

A drawback of the simple optical implementation of the technique presented here is that the pupil images will be blurred due to diffraction by the lenslet array. The finer the sampling in the sky image the worse the image quality in the pupil image. This effect prevents the DPI technique, as described here, from achieving correction at the diffraction limit of the full telescope aperture. Nevertheless, substantial correction to the natural seeing can be obtained, especially for very large telescope apertures. Table 1 shows how the spatial sampling obtained scales with telescope aperture at 500nm for pupil images on the detector with 10 resolution elements (limited by diffraction in the lens array) across the pupil. The relation is

$$\text{Sampling (mas)} = 2475/\text{aperture (m)}$$

| Aperture (m) | 4 | 8 | 30 | 100 |
|---|---|---|---|---|
| Scale (mas) | 619 | 309 | 82 | 25 |

**Table 1:** The relation between image (sky) plane sampling and telescope aperture at 500nm for a fixed number (10) of resolution elements across the individual pupil plane images.

## 6. CONCLUSIONS

There is a possible alternative to traditional AO systems for producing images with a spatial resolution substantially better than seeing-limited images on ELTs. This technique, which we call Dual Plane Imaging, is essentially a software based technique (in contrast to the hardware-based AO system approach).

Further work has to be done to investigate how well DPI would perform in practice. Three key areas need to be investigated:

- The feasibility of the detectors and the data capture system.
- The development of successful transformation algorithms.
- The development of an optical system which does not lose *directional* information due to diffraction effects.

## 7. ACKNOWLEDGEMENTS

We'd like to thank Marcel Carbillet (INAF:Arcetri) for access to the CAOS software. Anna Moore thanks the IoA, Cambridge, for financial support through their PPARC visitors program. She also thanks Sian Owen for providing excellent accommodation during her one month stay in Cambridge.